\documentclass[runningheads]{llncs}
\usepackage{bbding}
\usepackage[T1]{fontenc}
\usepackage{graphicx}
\usepackage{xurl}
\usepackage{hyperref}
\usepackage{color}

\begin{document}
\title{Mechanistic Interpretability Tool for AI Weather Models}
%
%
\author{\Envelope Kirsten I. Tempest\inst{1}\orcidID{0000-0002-2318-9032} \and
Matthias Beylich\inst{1}\orcidID{0009-0007-1734-2663} \and
George C. Craig\inst{1,2}\orcidID{0000-0002-7431-8164}}
\authorrunning{K.I. Tempest et al.}
%
\institute{Meteorological Institute Munich, Ludwig-Maximilians-Universität, 80333 Munich, Germany \and
Deutsches Zentrum für Luft- und Raumfahrt, Oberpfaffenhofen, Germany
\\
\email{k.tempest@physik.uni-muenchen.de}}

\maketitle              
\begin{abstract}
Artificial Intelligence (AI) weather models are improving rapidly, and their forecasts are already competitive with long-established traditional Numerical Weather Prediction (NWP). To build confidence in this new methodology, it is critical that we understand how these predictions are generated. This is a huge challenge as these AI weather models remain largely black boxes. In other areas of Machine Learning (ML), mechanistic interpretability has emerged as a framework for understanding ML predictions by analysing the building blocks responsible for them. Here we present an open-source, highly adaptable tool which incorporates concepts from mechanistic interpretability. The tool organises internal latent representations from the model processor and allows for initial analyses, including cosine similarity and Principal Component Analysis (PCA), enabling the user to identify directions in latent space potentially associated with meteorological features. Applying our tool to the graph neural network GraphCast, we present preliminary case studies for mid-latitude synoptic-scale waves and specific humidity. These demonstrate the tool's ability to identify linear combinations of latent channels that appear to correspond to interpretable features.

\end{abstract}
\section{Introduction}

Data-driven global weather forecast models developed in recent years are already competitive with traditional NWP, which has decades of research and development behind it. A major milestone was reached when AI models~\cite{Lam2023,bi2023} surpassed the forecast skill of the world-leading European Centre for Medium-Range Weather Forecasts (ECMWF) High RESolution forecast (HRES)~\cite{HRES}. Recent developments include foundation models such as Aurora~\cite{Bodnar2025}, and models designed for probabilistic forecasting such as AIFS-CRPS~\cite{lang2024}. The improvements in skill over time are tracked in the WeatherBench database~\cite{Rasp2024}.

The architectures of AI models and their resulting predictive capabilities are diverse. For example, GraphCast employs a graph neural network to produce a deterministic forecast, whereas AIFS-CRPS, a transformer-based model, generates ensemble members that sample a probability distribution. Despite these differences, these models share a common architectural framework. This consists of an encoding stage, a processor stage comprising multiple steps, including blocks that facilitate information exchange between locations, and a multilayer perceptron (MLP) operating on the latent space at each location, and finally a decoder stage which maps the learned representation back into gridded meteorological variables for humans to understand. This is quite different from conventional numerical integration, which evolves the physical variables through a sequence of time-steps to produce forecasts.
 
Traditional NWP systems have large and complex codebases, which are built on our understanding of physical processes in the atmosphere, as well as in other Earth system components. AI weather models on the other hand are optimised to reproduce a training dataset, without regard to prior physical knowledge. By iteratively adjusting the weights (often millions, if not a billion as in Aurora~\cite{Bodnar2025}) of the model during the training process, algorithms are created which allow the model to interpret input data and generate a meaningful output. This learning process allows the model to exploit any relevant relationships present in the weather data to be used in the production of the forecast, not just those programmed in. The trained model might make use of connections and concepts which are not yet understood or discovered by humans. This raises the question: how can we understand the internal workings of AI weather models?
 
There have been various studies analysing AI weather models. In a recent survey~\cite{Yang2024}, these approaches have been grouped into post-hoc interpretability techniques, and those where the design of the model is inherently interpretable. The former, relevant for understanding a broad range of models without altering the model itself, includes perturbation-based methods, where the input features are systematically modified and the predictions examined. Game theory based and gradient-based attribution methods have also been attempted. Although these methods provide insight, they remain specific to scenarios and models; a comprehensive understanding of how predictions are made, and what connections are learned, is missing. This is vital to build trust for integration of such models into operational weather prediction, and might lead to new scientific discoveries. 

Mechanistic interpretability seeks to reverse engineer neural networks by understanding interpretable features and the circuits that connect them~\cite{Olah2022}. Features, which are meaningful patterns in the data, can be identified to correspond with directions in the latent space. By identifying features at different stages in the processor, it is possible to learn algorithms in the weights of the model which are used to get to the final output. For example, a vision model classifies an image of a dog's head correctly by identifying eyes, snout, fur and tongue~\cite{Olah2020}. Mechanistic interpretability has been used as a framework in fields such as Large Language Models~\cite{Zhao2024,Kissane2024}, but has not yet been applied extensively to AI weather models. Initial work~\cite{MacMillan2025} has used sparse autoencoders to identify features in GraphCast, and has shown examples including surface heating in arid regions corresponding to directions in the latent space. This shows the potential of mechanistic interpretability techniques for understanding AI weather models.

Here we present an exploratory visualisation tool that leverages mechanistic interpretability concepts to investigate the latent space of AI weather models. By providing a structured interface, the tool enables millions of data points to be efficiently organised and interpreted, facilitating rapid idea and hypothesis generation. The tool is open-source and has been designed for ease of customisation, allowing users to extend supported model configurations and incorporate additional analyses. 

Section 2 provides an overview of the tool design and methods, including the model and data used. Section 3 presents two case studies, one focusing on mid-latitude synoptic-scale waves and the other on specific humidity and how they correspond to directions in the latent space. The paper is then concluded in Section 4. 

\section{Design and Method Overview}
\subsection{Design}

We propose a visualisation tool that can be used by both meteorologists and computer scientists to identify complex connection pathways within AI weather models. This requires the structured organisation and analysis of millions of data points.

In order to identify meaningful directions in the latent space, we begin by selecting a geographical region, one that contains a meteorologically relevant feature. Latent feature vectors within this region are then analysed. 

The tool comprises the following components: 

\begin{enumerate}
    \item Specify initial parameters. 
        \begin{itemize}
            \item Select the model configuration and forecast time (\(t\)). 
            \item Choose whether to apply the latent space translator. 
            \item Select \(T\), the number of maximally activated latent channels to display and use in analysis. 
        \end{itemize}

    \item Select location.
        \begin{itemize}
            \item Choose the meteorological variable to be displayed on the two maps. The first map shows the input data at the forecast initialisation time, \(f(t_{\mathrm{init}}\)), and the second shows the increment of the atmospheric state between initialisation and forecast time, \(f(t) - f(t_{\mathrm{init}}\)). 
            \item Define a circular geographic region of interest.
        \end{itemize}

    \item Extract latent feature vectors and perform initial analysis.
        \begin{itemize}
            \item Latent feature vectors are loaded for all processor steps and mesh nodes, and the mesh nodes within the selected area are identified.
            \item The \(T\) most strongly activated latent channels are selected, and global maps of their activations are plotted for the selected processor step, \(P\).
        \end{itemize}

    \item Perform further analysis.
        \begin{itemize}
            \item Cosine similarity analysis.
            \item Principal Component Analysis (PCA), with user-specified number of components, \(C\).
        \end{itemize}
\end{enumerate}

We use Streamlit~\cite{Streamlit}, an open-source app framework available as a Python package. It avoids a backend due to being able to add widgets in the same manner as declaring a variable. The aim is for scientists to use this open-source tool, provide feedback, or develop their own versions by forking the GitHub repository and customising it with their own data and analysis methods. We hope this allows for fast development in the field of AI weather model interpretability.

\subsection{GraphCast}

To demonstrate the visualisation tool, we will examine the GraphCast model~\cite{Lam2023}. This model is capable of producing high-quality forecasts while being relatively compact in size. The input reanalysis dataset it is trained on, ERA5~\cite{Hersbach2020}, is open-source, and can therefore aid in investigating connections between directions and features. 

For our initial analysis, we use the pretrained small version of GraphCast, with 1 degree resolution. However, the tool will function for other configurations of GraphCast, and can be extended for use with other AI weather models. 

As we focus exclusively on latent representations, the model is run for a single time step (6 hours in the case of GraphCast). Consequently, the time difference between initialisation and forecast is fixed at 6 hours.

\subsubsection{Processor Step}

The processor stage of GraphCast is made up of 16 processor steps, each consisting of a deep graph neural network operating on a multi-mesh, where nodes are connected by edges of varying lengths. The multi-mesh is constructed by combining icosahedral meshes of differing resolution, from the base mesh with 12 nodes to the finest resolution which has 40,962 nodes. In the small version used in this paper, the finest resolution contains 10,242 nodes.

Within each processor step, edge representations are first updated using information from adjacent mesh nodes. The mesh nodes are then updated by aggregating information from incoming edges. The resulting node representations are then normalised and updated via residual connections. This message-passing procedure is repeated iteratively.

\subsubsection{Latent Feature Vectors}

We are interested in the latent feature vectors at each mesh node across the globe, extracted at the end of each processor step. These are of length 512, and whose individual entries we refer to as channels. For the small version of GraphCast, there are 10,242 mesh nodes at each processor step, with feature vectors of length 512. As such, there are a total of 83,902,464 latent data points to visualise and understand. 

Meteorologically interpretable feature vectors may be arbitrary linear combinations of the channels in the latent space, and their orientation may be transformed between different processor steps. To compare the latent feature vectors consistently across these steps, the tool provides an option to apply an affine transformation, referred to as a translator, to the latent feature vectors~\cite{Beylich2026}. Whenever intermediate processor steps are analysed in this paper, the translator is applied so that the basis is the same as that of the final processor step.

\subsection{Analysis}

\subsubsection{Cosine Similarity}
Cosine Similarity is the normalised dot product of two vectors, \(\bar A\) and \(\bar B\)~\cite{scikit-learn_cosine_similarity}, 
\noindent
\begin{equation}
\frac{\bar A\cdot \bar B}{\|\bar A\|\|\bar B\|}.
\end{equation}

It is bounded between \(-1\) and \(1\), whereby a value of \(1\) means that the vectors are perfectly aligned, \(-1\) indicates that they are oppositely aligned, and \(0\) denotes that they are orthogonal and therefore not similar at all.

In this analysis, the channels with the highest activations are selected to create a vector of length \(T\) at every mesh node. A node is then chosen at random from within the selected region, and the vector at that node is compared will all other mesh nodes globally. The analysis is then repeated, but comparing the full latent feature vector rather than shortening it to the most activated channels. 

\subsubsection{Principal Component Analysis (PCA)} 
PCA is used to project data onto a lower-dimensional space defined by \(C\) principal components that capture the greatest variance~\cite{scikit-learn_PCA}. This enables the most prominent features to be found.

In this analysis, the data in the selected geographical region is used to fit the principal components, and the global data is then transformed into this new basis. The amount of variance each principal component captures is calculated, as well as the channels that contribute most to each principal component vector. 

\section{Results: Case Studies}

Two case studies are presented to illustrate the capabilities of the tool for interpreting AI weather models. In each case, a geographical region containing a relevant meteorological feature is identified, along with the latent channels maximally activated within that region. Global maps of these channel activations, as well as the resulting cosine similarity and PCA analyses, are then visualised to assess whether possible connections can be identified between the latent feature vectors and meteorological features.

As our primary aim is to identify directions associated with such features, we focus on the final (\(16^{\mathrm{th}}\)) processor step unless otherwise stated. Furthermore, we use the top 15 maximally activated latent channels \(T\), and for PCA, the \(C=8\) leading components are extracted.

\subsection{Mid-latitude Synoptic-Scale Waves}

\begin{figure}[h]
\includegraphics[width=\textwidth]{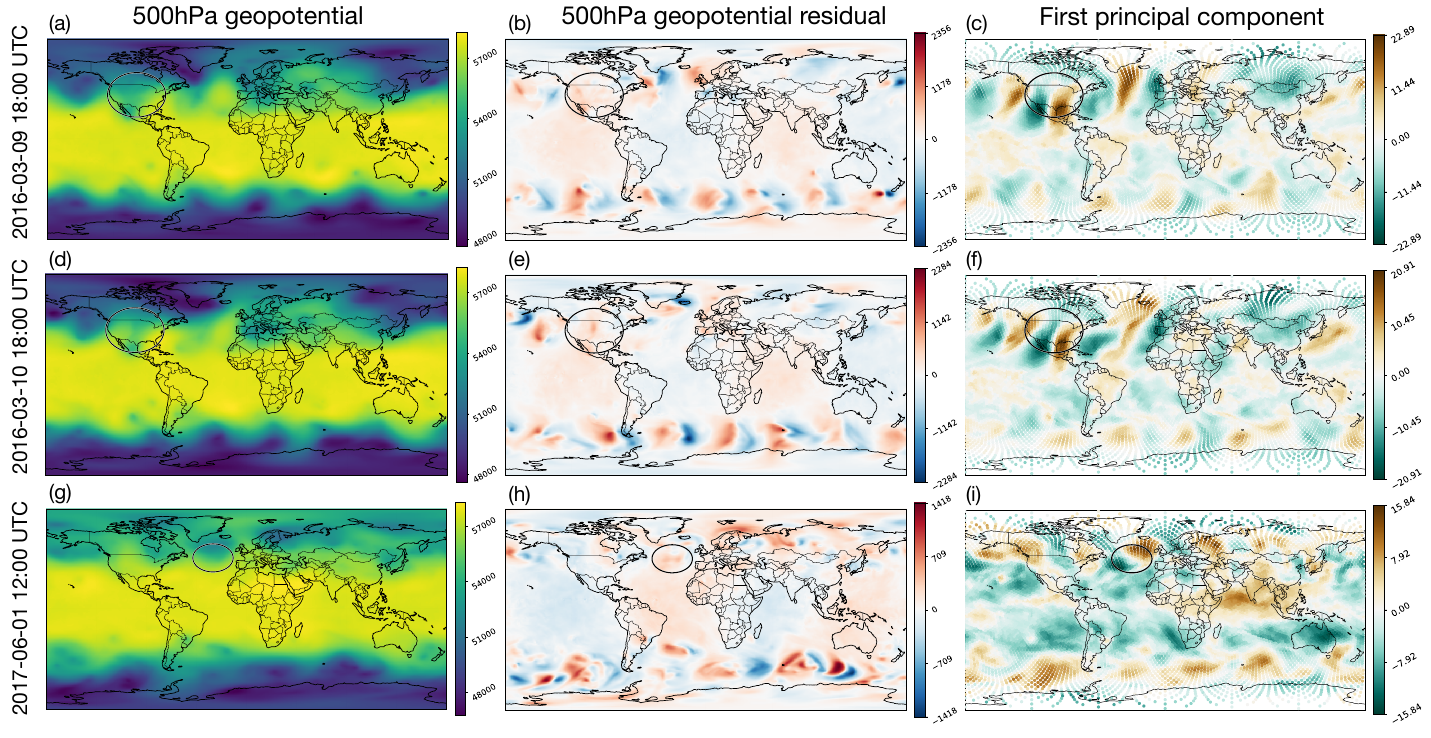}
\caption{Global fields for three forecast times, \(t\): (a-c) 2016-03-09 18:00 UTC, (d-f) 2016-03-10 18:00 UTC, and  (g-i) 2017-06-01 12:00 UTC. The left column (a,d,g) shows the 500hPa geopotential (\(m^2 s^{-2}\)) at the forecast initialisation time, \(t_{\mathrm{init}}\). The middle column (b,e,h) shows the corresponding residual (\(f(t) - f(t_{\mathrm{init}}\))), and the right column (c,f,i) the first principal component. The circled region highlights the area used for analysis. For the first two rows, the circle is centred at \(40^{\circ}N\), \(100^{\circ}W\) and has a radius of \(20^{\circ}\). In the bottom row, the circle is centred at \(46^{\circ}N\), \(30^{\circ}W\) and has a radius of \(12.59^{\circ}\).} \label{fig1}
\end{figure}

We first identify directions associated with synoptic-scale waves in the mid-latitudes. A trough of low pressure is selected at \(40^{\circ}N\), \(100^{\circ}W\), indicated by the black circle with radius \(20^{\circ}\) in Fig.~\ref{fig1}. The initial forecast time analysed is 2016-03-09 at 18:00 UTC, and the last ERA5 input (initial condition) is 6 hours before, at 12:00 UTC (Fig.~\ref{fig1}a). The residual (Fig.~\ref{fig1}b) represents the 6~h forecast increment generated by the AI model. 

The first principal component in Fig.~\ref{fig1}c shows a pronounced alternating dipole structure in the Northern mid-latitudes, particularly in the West. This suggests that a dipole (negative activation to the west, positive activation to the east) is associated with each trough. Focusing on the two troughs nearest the western edge of the domain, it is seen that the eastern trough extends further south. Similarly, the dipoles extend further south. Although the dipoles are weaker in the east, the connection is still evident. Dipole structures are also present in the Southern Hemisphere, but are considerably weaker than that observed in the Northern Hemisphere. 

To investigate how the connection evolves over time, Fig.~\ref{fig1}d-f shows the same fields 24 hours later. This period is associated with a significant blocking event~\cite{Hauser2023}, which slows the eastward propagation of waves, with movement occurring primarily south of Greenland. It is observed that the low geopotential has shifted eastward there, accompanied by a corresponding shift in the dipole structure.

The last row of Fig.~\ref{fig1} shows a forecast from a different time of year (2017-06-01 12:00 UTC) to assess whether the dipole structure in the principal component persists when the synoptic-scale waves are not as clearly defined. At a trough (of radius \(12.59^{\circ}\)) at \(46^{\circ}N\), \(30^{\circ}W\), the same dipole structure is observed, but tilted in the opposite direction to the previous case. Furthermore, the strong alternating dipole structure across the Northern Hemisphere is less pronounced. 

\begin{figure}
\includegraphics[width=\textwidth]{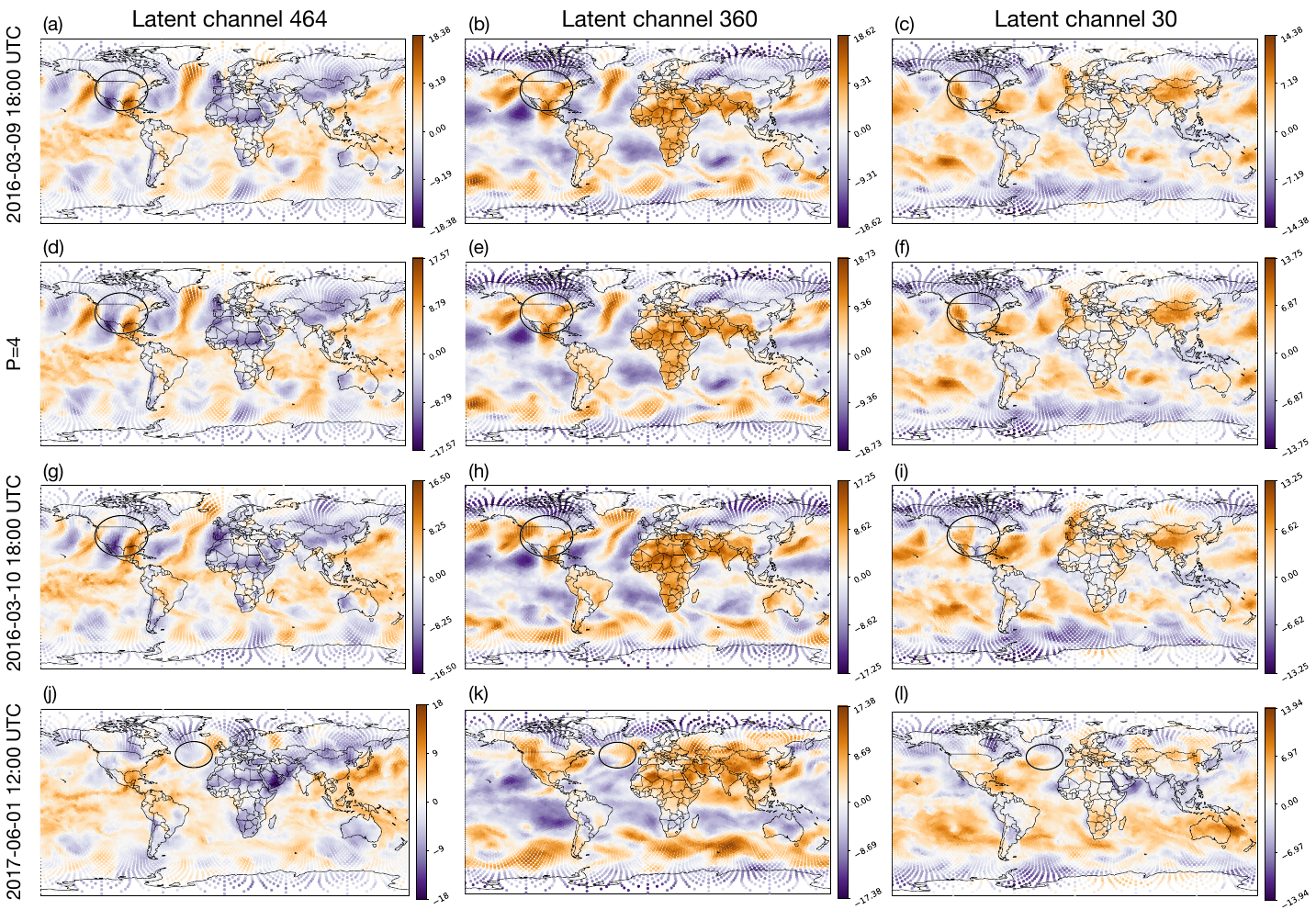}
\caption{Spatial structure of selected latent channels for three forecast times. Rows (a–c), (g–i) and (j–l) show the activation of channels 464, 360 and 30, respectively, at the last (\(16^\mathrm{th}\)) processor step for forecast times 2016-03-09 18:00 UTC, 2016-03-10 18:00 UTC and 2017-06-01 12:00 UTC. Second row from top (d–f) corresponds to the same forecast time as (a–c), but shows an earlier processor step (\(P=4\)). Circles correspond to the same locations as in Fig.~\ref{fig1}.} \label{fig2}
\end{figure}

The top three channels contributing to the first principal component at the first forecast time of 2016-03-09 18:00 UTC are shown in Fig.~\ref{fig2}, and Table~\ref{tab1} details the top contributing channels for all forecast times from Fig.~\ref{fig1}. There is general consistency in the channels contributing to the first principal component across forecast times, although the ranking is not identical. When a distinct trough is present in the geopotential, the dipole structure is prominent in both channels 464 and 360, and, to a lesser extent, in channel 30, where the activation signs are reversed. Additional patterns are also seen in the raw channels. For example, channel 360 is often strongly activated over most of Africa. Finally, we look at the latent channels at an earlier processor step. Fig.~\ref{fig2}d-f shows the latent channels at processor step four for the forecast time of 2016-03-09 18:00 UTC. There are only minute differences; the earlier processor step already shows the broad structure observed in the final processor step. 

\begin{table}
\caption{Top channels contributing to the first principal component at the final processor step, \(P=16\), for the synoptic wave case. Descending order, with \(1\) contributing most. Channels in bold occur at least twice.}\label{tab1}
\begin{tabular}{|l|c|c|c|c|c|c|}
\hline
Date and time of forecast &  1 & 2 & 3 & 4 & 5 & 6 \\
\hline
2016-03-09 18:00 UTC & \textbf{464} & \textbf{360} & \textbf{33} & 239 & \textbf{269} & \textbf{30} \\
2016-03-10 18:00 UTC & \textbf{464} & \textbf{360} & \textbf{33} & \textbf{126} & \textbf{30} & \textbf{269} \\
2017-06-01 12:00 UTC & \textbf{360} & 19 & \textbf{30} & \textbf{126} & 183 & 426 \\
\hline
\end{tabular}
\end{table}

\begin{figure}[h]
\includegraphics[width=\textwidth]{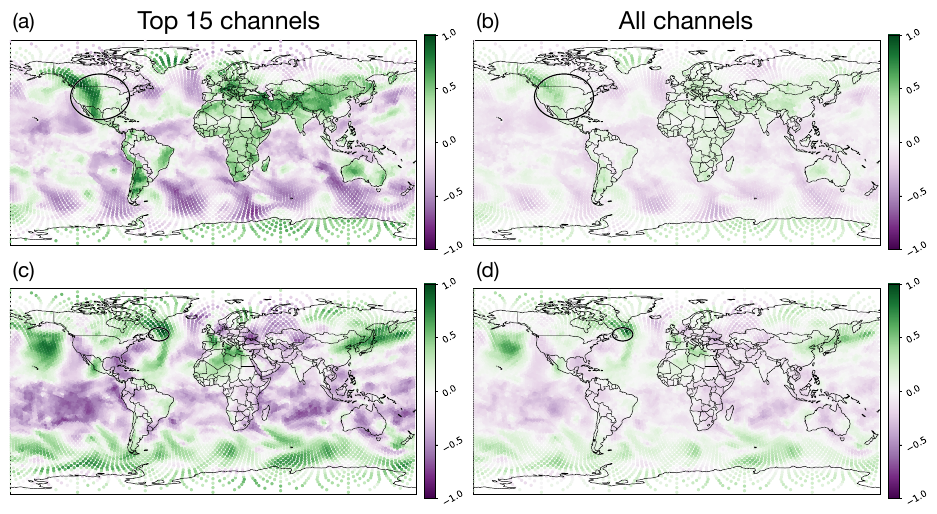}
\caption{Cosine similarity of the latent feature vectors for the forecast at 2016-03-09 18:00 UTC, evaluated using two regions indicated by black circles; (a-b) analysis region is as in top two rows of Fig.~\ref{fig1}, and (c-d) analysis region is centred at \(50^{\circ}N\), \(48^{\circ}W\) and has a radius of \(5.81^{\circ}\). Left column (a,c) shows similarity computed using the 15 most activated channels, whereas the right column (b,d), uses the full set of channels.} \label{fig3}
\end{figure}

The cosine similarity diagnostic is used to identify regions with similar latent space vectors. At first glance, the cosine similarity using the previously selected region over North America (Fig.~\ref{fig3}a,b) does not yield obviously interpretable patterns. However, when a different region in the mid-latitudes is selected from the same forecast, a pattern of synoptic-scale waves emerges (Fig.~\ref{fig3}c,d). Interestingly, this pattern does not include the North American trough, perhaps because geopotential values are not sufficiently low. When the entire vector is used for the cosine similarity as in Fig.~\ref{fig3}d, the same pattern as in Fig.~\ref{fig3}c emerges, but with reduced magnitude. 

\subsection{Specific Humidity}

\begin{figure}[h]
\includegraphics[width=\textwidth]{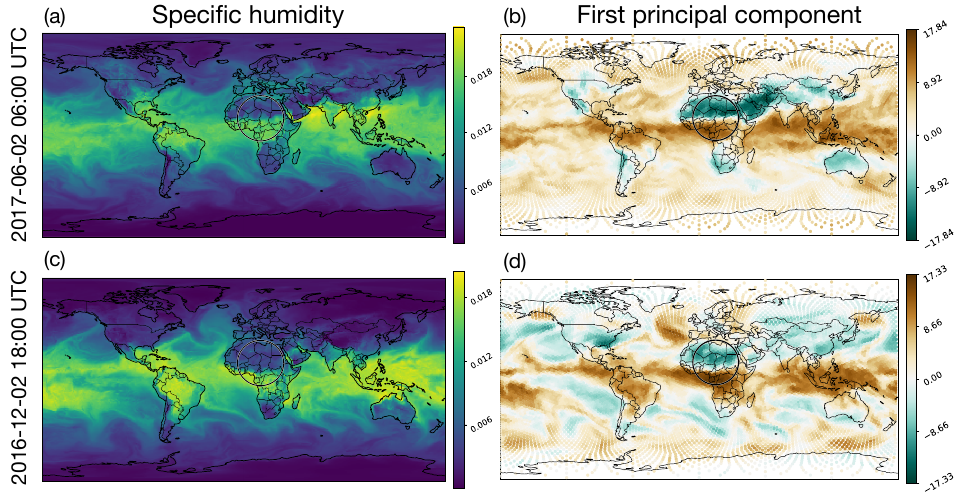}
\caption{Global fields for two forecast times, \(t\): (a-b) 2017-06-02 06:00 UTC and (c-d) 2016-12-02 18:00 UTC. The left column (a,c) shows the 1000hPa specific humidity (\(kg kg^{-1}\)) at the forecast initialisation time, \(t_{\mathrm{init}}\), and the right column (b,d), the first principal component. The circled region highlights the area used for analysis. For both rows, the circle is centred at \(15^{\circ}N\), \(15^{\circ}E\) and has a radius of \(20^{\circ}\).} \label{fig4}
\end{figure}

Next, we identify directions in the latent space associated with specific humidity. Fig.~\ref{fig4} shows the ERA5 input data for specific humidity (a,c) and the first principal component (b,d) for two dates and times, one in winter and one in summer (rows). A region of radius \(20^{\circ}\) is selected, centred at \(15^{\circ}N\), \(15^{\circ}E\), which includes the strong moisture gradient of the Sahel region of Africa.

A clear correlation between the first principal component and the specific humidity is observed. This is evident north of the Gulf of Guinea. For example, on 2017-06-02 at 06:00 UTC, there is a relatively high specific humidity of approximately 0.015 \(kg/kg\) reaching up to the Northern borders of Nigeria. In contrast, on 2016-12-02 at 18:00 UTC, those levels of specific humidity extend only half way up the country. Similarly, a strong activation in the principal component is seen for the full extent of Nigeria on 2017-06-02 and only half way up on 2016-12-02. A clear correlation is also seen over Australia. The correspondence between the principal component and the meteorological data is also visible over the oceans. For example, the swirl on 2016-12-02 in the Atlantic is captured, as well as the prong-like structure in the Eastern Pacific. These correlations are strongest in the lower latitudes, weakening at higher latitudes, around \(50^{\circ}\).

\begin{figure}
\includegraphics[width=\textwidth]{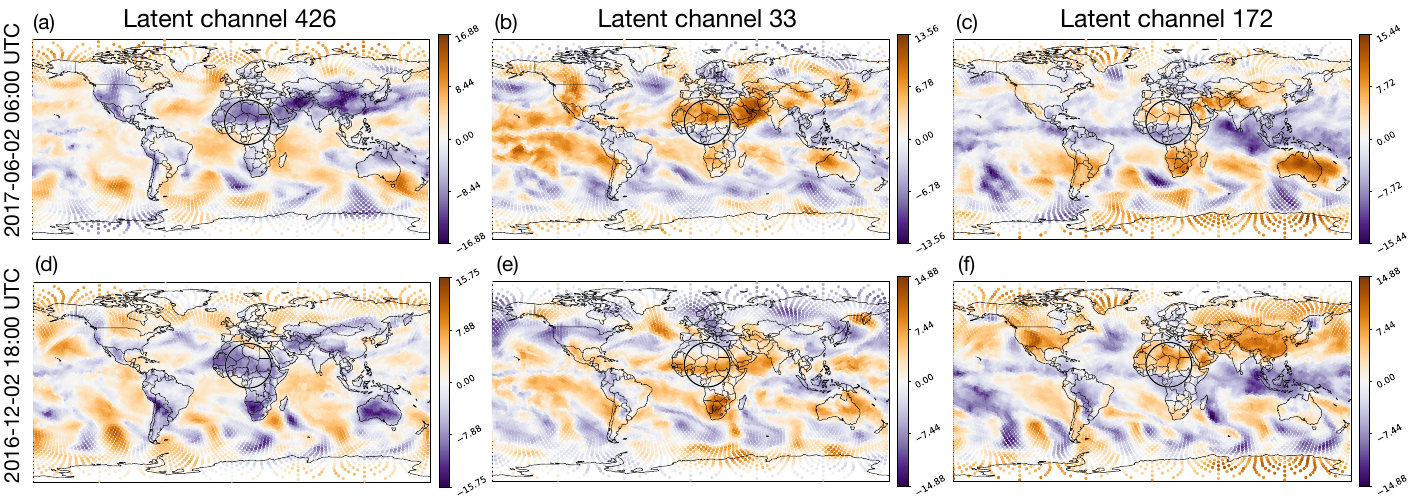}
\caption{Spatial structure of selected latent channels for two forecast times. Rows (a–c) and (d–f) show the activation of channels 426, 33 and 172, respectively, at the last (\(16^\mathrm{th}\)) processor step for forecast times 2017-06-02 06:00 UTC and 2016-12-02 18:00 UTC. Circles correspond to the same location as in Fig.~\ref{fig4}.} \label{fig5}
\end{figure}

As before, the channels contributing to the first principal component are detailed in Table~\ref{tab2}, and the top three from the first row are shown in Fig.~\ref{fig5} for the two forecast dates. Once again, there is a noticeable consistency over the two times, with all but one of the top six channels appearing in both forecasts, although not in the same order. Apart from a clear break at the equator and a strong activation either side of this in channels 33 and 172, distinct spatial patterns are less evident in the individual channels than for the synoptic wave feature. It is furthermore worth noting that a couple of channels appearing in Table~\ref{tab1} also occur in Table~\ref{tab2}.

\begin{table}
\caption{Top channels contributing to the first principal component at the final processor step, \(P=16\), for the specific humidity case. Descending order, with \(1\) contributing most. Channels in bold occur twice.}\label{tab2}
\begin{tabular}{|l|c|c|c|c|c|c|}
\hline
Date and time of forecast &  1 & 2 & 3 & 4 & 5 & 6 \\
\hline
2017-06-02 06:00 UTC & \textbf{426} & \textbf{172} & 33 & \textbf{464} & \textbf{19} & \textbf{183} \\
2016-12-02 18:00 UTC & \textbf{183} & \textbf{172} & \textbf{464} & \textbf{19} & \textbf{426} & 360 \\
\hline
\end{tabular}
\end{table}

\subsection{Discussion}

From the two case studies, we have identified possible connections between directions in the latent space and meteorological features. This illustrates the usefulness of the tool for exploring the latent space of the processor stage in AI weather models and generating hypotheses for further investigation.

Across the case studies, some components of the tool were more informative than others. In particular, PCA was effective at identifying prominent structures in the latent space and highlighting the channels that contributed most strongly to them. In both cases, each of the two features studied were associated with a set of frequently identified channels, although not always in the same order. Furthermore, although the two features we studied are not directly related, and are of importance in different geographical regions, some channels contributed strongly to both features e.g. channels 360 and 464. This might suggest a physical relationship between specific humidity and mid-latitude synoptic-scale waves, since they are represented in part by the same underlying features, although it could also indicate polysemantic neurons.

The ability to quickly generate visualisations of the latent space for different regions and then relate them to the atmospheric state resulted in the two case studies presented here, and suggests several avenues for further investigation using more quantitative techniques. For instance, our initial results suggest that AI weather models may represent the Northern and Southern Hemispheres differently, which may reflect differences in land-sea distribution. Although parts of some channels could be interpreted, other activated regions remained difficult to explain. For example, in channel 360, a recurring dipole activation was observed in the mid-latitudes, while strong activation over Africa remained unexplained.

Another application of the visualisation tool that has not been explored extensively here is how the latent space evolves across processor steps. In the one example shown in Fig.~\ref{fig2}, it was observed that differences between steps four and sixteen were minimal, suggesting that certain large-scale features may already be established at earlier processor steps.

\section{Conclusion}

We have presented a tool to support mechanistic interpretability research in AI weather models, which currently supports the graph neural network GraphCast. The tool enables systematic organisation and visualisation of large-scale meteorological and latent data, while also supporting initial analyses. It is open source and available online at \href{https://github.com/ktempestuous/latent_space_visualiser_weather_models}{GitHub}. 

The aim of the tool is to enable rapid visual exploration of the model's latent space, to generate hypotheses that can motivate later, more detailed analysis using quantitative methods on large data sets.
The use of the tool has been demonstrated in two case studies. In both cases, qualitative associations were observed between directions in the latent space and meteorological features, specifically mid-latitude synoptic-scale waves and specific humidity. These results highlight the potential to identify further feature–direction correspondences and to develop a deeper understanding of those already identified. In turn, this could enable the construction of a dictionary of interpretable latent features. Which could then support the analysis of how feature-associated directions evolve across processor steps and lead to the identification of circuits, ultimately contributing to a more comprehensive understanding of how AI weather models generate their predictions.

Development of the tool is ongoing, with current work focused on accommodating transformer-based models, and on supporting visualisations across selected forecast times.

\begin{credits}

\subsubsection{Code and Data Availability}
The visualisation tool developed in this work is open source and available at \url{https://github.com/ktempestuous/latent_space_visualiser_weather_models}. The repository includes instructions for installation, configuration, and extension to additional model architectures. In addition, sample data is accessible to use with the tool.

The ERA5 reanalysis dataset used in this study is publicly available from ECMWF at \url{https://www.ecmwf.int/en/forecasts/datasets/}.

Latent feature datasets were created by running the GraphCast model available at \url{https://github.com/google-deepmind/graphcast} with the small configuration, and extracting the latent features after each processor step.

\subsubsection{\ackname} The authors thank the computing resources and support of the Physics department at LMU, as well as fruitful discussions with colleagues for advancing this topic. 

\subsubsection{\discintname}
The authors have no competing interests to declare that are
relevant to the content of this article.
\end{credits}

%
%
%
\bibliographystyle{splncs04}
\bibliography{references}

\end{document}